\documentclass[sigconf, nonacm]{acmart}
\usepackage{graphicx} 
\usepackage{listings}
\usepackage{xcolor}
\usepackage{multirow}
\usepackage{algorithm}
\usepackage{algpseudocode}
\usepackage{amsmath}
\usepackage{balance}
\usepackage{tikz}
\usepackage{fontawesome5}
\usepackage{footmisc}
\usepackage{tcolorbox}
\usetikzlibrary{shapes.geometric, arrows.meta, positioning, backgrounds, fit, shadows}
\usepackage{etoolbox}
\usepackage{booktabs}
\usepackage{paralist}
\usepackage{enumitem}
\setlist[itemize]{noitemsep, topsep=0pt}
\makeatletter
\algrenewcommand\algorithmiccomment[1]{%
  \hfill\(\triangleright\)\,{\footnotesize\ttfamily\textcolor{gray}{#1}}%
}
\makeatother

\tikzstyle{tool} = [draw=gray, fill=gray!20, rounded corners, minimum width=4.0cm, minimum height=0.5cm, font=\normalsize]
\tikzstyle{process} = [draw=gray, fill=gray!20, rounded corners, minimum width=4.5cm, minimum height=0.5cm, font=\normalsize]
\tikzstyle{manual} = [draw=gray, fill=gray!20, rounded corners, minimum width=4.5cm, minimum height=0.5cm, font=\normalsize]
\tikzstyle{output} = [draw=gray, fill=gray!20, rounded corners, minimum width=4.5cm, minimum height=0.5cm, font=\normalsize]
\tikzstyle{arrow} = [thick,->,>=Stealth]

\AtBeginDocument{%
  }

\setcopyright{none}
\copyrightyear{2026}
\acmYear{2026}
\acmDOI{XXXXXXX.XXXXXXX}

\begin{document}

\title{HistoryFinder: Advancing Method-Level Source Code History Generation with Accurate Oracles and Enhanced Algorithm}


\author{Shahidul Islam}
\email{islams32@myumanitoba.ca}
\affiliation{%
  \institution{SQM Research Lab, University of Manitoba}
  \city{Winnipeg}
  \state{Manitoba}
  \country{Canada}
}

\author{Ashik Aowal}
\email{r.aowal@gmail.com}
\affiliation{%
  \institution{American International University-Bangladesh}
  \city{Dhaka}
  \state{Dhaka}
  \country{Bangladesh}
}

\author{Md Sharif Uddin}
\email{s.uddin@usask.ca}
\affiliation{%
  \institution{University of Saskatchewan}
  \city{Saskatoon}
  \state{Saskatchewan}
  \country{Canada}
}

\author{Shaiful Chowdhury}
\email{shaiful.chowdhury@umanitoba.ca}
\affiliation{%
  \institution{SQM Research Lab, University of Manitoba}
  \city{Winnipeg}
  \state{MB}
  \country{Canada}
}




\begin{abstract}
Reconstructing a method’s change history efficiently and accurately is critical for many software engineering tasks, including maintenance, refactoring, and comprehension. Despite the availability of method history generation tools such as \textit{CodeShovel} and \textit{CodeTracker}, existing evaluations of their effectiveness are limited by inaccuracies in the ground truth oracles used. In this study, we systematically construct two new oracles—the corrected \textit{CodeShovel} oracle and a newly developed \emph{HistoryFinder oracle}—by combining automated analysis with expert-guided manual validation. We also introduce \textit{HistoryFinder}, a new method history generation tool designed to improve not only the accuracy and completeness of method change histories but also to offer competitive runtime performance. Through extensive evaluation across 400 methods from 40 open-source repositories, we show that \textit{HistoryFinder} consistently outperforms \textit{CodeShovel}, \textit{CodeTracker}, \textit{IntelliJ}, and \texttt{Git-based} baselines in terms of precision, recall, and F1 score. Moreover, \textit{HistoryFinder} achieves competitive runtime performance, offering the lowest mean and median execution times among all the research-based tools. While \texttt{Git-based} tools exhibit the fastest runtimes, this efficiency comes at the cost of significantly lower precision and recall—leaving \textit{HistoryFinder} as the best overall choice when both accuracy and efficiency are important. To facilitate adoption, we provide a web interface, CLI, and Java library for flexible usage.
\end{abstract}



\keywords{method evolution, software evolution, mining software repositories}



\maketitle
\newcommand{\CodeShovelDisplayErrorsFootnote}{
  Missing \texttt{displayErrors} method change in CodeShovel: 
  \href{https://github.com/ataraxie/codeshovel/blob/30dbe880cb6766612ddf0ee406c10e1461c902f7/src/test/resources/oracles/java/checkstyle-Checker-fireErrors.json}{https://github.com/ataraxie/codeshovel/blob/..}
}

\newcommand{\CodeTrackerDisplayErrorsFootnote}{
  Missing \texttt{displayErrors} method change in CodeTracker: 
  \href{https://github.com/jodavimehran/code-tracker/blob/f9b4f6c5055ceeb8ab781168c6db0cccd2ca0c85/src/main/resources/oracle/method/training/checkstyle-Checker-fireErrors.json}{https://github.com/jodavimehran/code-tracker/blob/..}
}

\newcommand{\MethodRenameFootnote}{
  Method rename diff from \texttt{displayErrors} to \texttt{fireErrors}: 
  \href{https://github.com/checkstyle/checkstyle/compare/f34eba11cbd8559d4d06f96e19affa70abaf86ff...0e3fe5643667a53079dbd114e5b1e9aa91fde083\#diff-b48cc2ff38026f94180847550cd260a0ebefb4a9cb9635e39f5fa2046311d72b}{https://github.com/checkstyle/checkstyle/compare/..}
}

\newcommand{\CodeShovelOverloadedCreatePatternFootnote}{
CodeShovel misidentifies \texttt{createPattern}:
  \href{https://github.com/checkstyle/checkstyle/compare/d2551035044a845fd1b3e345f2470875a43a8991...f2c6263e151e8a7300755b36fbb41511c0a0cca1\#diff-1b66e778c19cac5ac2beaec4d6a74c8c37a5fd83d39db4bef8c1b60b9ec68c0e}{https://github.com/../d25510..}
}

\newcommand{\CodeTrackerMergeCommitRunChildFootnote}{
  Incorrect \texttt{runChild} merge diff in CodeTracker:
  \href{https://github.com/junit-team/junit4/compare/8ff0b44e3fb0c1c56a1dc6290c3b6828a5a8f9bf...bed58a573c373d57d64fa369f58b2a8f0dbee3d7\#diff-cd3bb9170f8d6404c5dc5bb66f7b647435854d48fe357e0bd998e3011a575929}{https://github.com/../8ff0b4..}
}

\newcommand{\RunChildMergeCommitParentFootnote}{
  \texttt{runChild} method in merge commit: 
  \href{https://github.com/junit-team/junit4/blob/a49240ade1974b948b20cf2c45d9129f04122735/src/main/java/org/junit/runners/BlockJUnit4ClassRunner.java\#L64}{https://github.com/../junit4/blob/a49240..}
}

\newcommand{\RunChildWrongParentFootnote}{
  \texttt{runChild} method in wrong commit: 
  \href{https://github.com/junit-team/junit4/blob/708ed373c02b467422890d5163fff46d1f32ab01/src/main/java/org/junit/runners/BlockJUnit4ClassRunner.java\#L66}{https://github.com/../junit4/blob/708ed3..}
}

\newcommand{\RunChildCorrectParentFootnote}{
  \texttt{runChild} in method correct commit: 
  \href{https://github.com/junit-team/junit4/blob/9d8bb069f68e2194db742981972c8930381b62c2/src/main/java/org/junit/runners/BlockJUnit4ClassRunner.java\#L65}{https://github.com/../junit4/blob/9d8bb0..}
}

\section{Introduction}
Source code change history is valuable to both software practitioners and researchers. Practitioners leverage source code history for tasks such as code reviews, expert identification, and developer onboarding~\cite{grund_codeshovel_2021}. Researchers, on the other hand, use code history for a variety of tasks, including bug prediction~\cite{chowdhury_method-level_2024, mashhadi_empirical_2024}, code clone management~\cite{duala-ekoko_tracking_2007}, entity name recommendation~\cite{kashiwabara_method_2015}, security vulnerability detection \cite{li_cleanvul_2024}, code review \cite{grund_codeshovel_2021}, authorship attribution ~\cite{meng_mining_2013, rahman_ownership_2011}, and software maintenance effort estimation~\cite{chowdhury_good_2025, chowdhury_evidence_2024, hassan_versioned_2024}. To extract this historical information, researchers primarily relied on version control systems (VCS). However, traditional \texttt{VCSs} like \texttt{Git} mainly focus on file-level change tracking, which lacks the granularity needed for studies at finer levels. In practice, understanding maintenance activities at lower granularities—particularly at the method level—is important to both researchers and practitioners~\cite{grund_codeshovel_2021, chowdhury_evidence_2024, chowdhury_method-level_2024}. However, tracking method-level change history is challenging, as methods may be renamed or moved across files during refactoring and other development activities. These challenges and practical demands have driven the development of specialized tools for generating method-level change histories~\cite{higo_tracking_2020, grund_codeshovel_2021, jodavi_accurate_2022}.


Several approaches have been proposed to recover method-level code evolution, particularly from \texttt{Git} repositories. Hata \textit{et al.}~\cite{hata_historage_2011} introduced \textit{Historage}, a tool that enables method-level change tracking by restructuring repositories to store individual methods in separate files. This approach was later refined by \textit{FinerGit}~\cite{higo_tracking_2020}, which enhanced the accuracy of detecting small method changes. However, both tools require extensive preprocessing~\cite{grund_codeshovel_2021} of the codebase before any method-level history can be generated. This introduces additional storage overhead and processing time, making them impractical for scenarios where fast and on-demand change history extraction is needed.


To mitigate the need for upfront preprocessing, Grund \textit{et al.} ~\cite{grund_codeshovel_2021} introduced \textit{CodeShovel} (ICSE 2021), a tool that tracks method-level change history on demand using a lightweight string similarity algorithm. CodeShovel was reported to be both accurate and efficient, based on evaluation using a human-generated change history oracle of 200 Java methods. 
Unfortunately, Jodavi \textit{et al.}~\cite{jodavi_accurate_2022} later identified inaccuracies in this oracle. They corrected the oracle and introduced a more robust and refactoring-aware tool,  \textit{CodeTracker} (ESEC/FSE 2022), which significantly outperformed \textit{CodeShovel} in terms of precision and recall.

We are, however, concerned that the internal threats present in the human-generated \textit{CodeShovel} oracle may have carried over into the \textit{CodeTracker} oracle, which was largely derived from \textit{CodeShovel} and later corrected by a separate group of researchers—introducing its own set of internal threats. Such inaccuracies in recorded change histories are particularly problematic, as tools validated by these oracles form the foundation for research on method-level maintenance analysis and modeling~\cite{ahmad_impact_2025, chowdhury_evidence_2024, chowdhury_good_2025, chowdhury_method-level_2024, chowdhury_revisiting_2022}. Moreover, the accuracy of tools evaluated against these less reliable oracles remains uncertain, raising concerns about their practical usefulness to software practitioners. Low precision in identifying method change commits can result in excessive false positives, while low recall may lead to missing critical changes—both of which can harm the usefulness of change history tracking in real-world scenarios.

To address the limitations of human-generated change histories, we adopted a more systematic, semi-automated approach and constructed two new oracles for method-level change history. Rather than relying solely on manual annotation, we developed a tool that, for a given Java method, collects method change commits identified by multiple state-of-the-art tools—namely, \textit{CodeShovel}, \textit{CodeTracker}, \textit{GitLog} (\textit{GitFuncName}, and \textit{GitLineRange}), and \textit{IntelliJ}. The tool aggregates the union of method change commits reported by all sources, based on our observation that this combined approach achieves significantly higher recall than any individual tool. This union set was then manually verified by two expert software engineers, who removed false positives to ensure high precision and added any relevant method change commits missed by all tools. The result is two high-quality oracles that combine the strengths of automated detection and expert validation. As we show later in our analysis (\textbf{RQ1}), the previously used oracles from \textit{CodeShovel} and \textit{CodeTracker} contain notable inaccuracies—confirming our concerns about the validity of prior evaluations that relied on those oracles.  

In addition to providing the new oracles, we developed \textit{HistoryFinder}, a novel, heuristic-based, method-level history generation algorithm and tool. \textit{HistoryFinder} also addresses several of \textit{CodeShovel}'s technical limitations. For instance, unlike \textit{CodeShovel}, \textit{HistoryFinder} captures changes introduced in merge commits, as well as modifications to Javadocs, annotations, and code formatting.

We then evaluated all method-level change history tracking tools, including \textit{HistoryFinder}, in terms of precision, recall, and runtime when generating change histories for the two newly developed oracles (\textbf{RQ2} and \textbf{RQ3}). Overall, the research-based tools—\textit{HistoryFinder}, \textit{CodeShovel}, and \textit{CodeTracker}—consistently demonstrate substantially higher accuracy than commonly used alternatives such as \textit{GitLog} and \textit{IntelliJ}. \textit{HistoryFinder} generally achieves the highest precision and recall across all oracles, while also offering competitive runtime performance. Although \textit{GitLog} is the fastest at reconstructing method histories, it suffers from significantly lower precision and recall, making \textit{HistoryFinder} the most balanced option in terms of accuracy and efficiency. To support broader adoption, we also provide a \texttt{GUI-based} tool that allows researchers and practitioners to generate method-level change histories using any of the state-of-the-art tools.

The contribution of our paper can be summarized as follows.
\begin{itemize}
    \item We provide two systematically developed method-level change history oracles---consisting of 400 Java methods. Building these oracles with manual verification required approximately 300 hours from each of the two participating authors, excluding the time for building the necessary tools. These more accurate new oracles can be instrumental for future research on developing more method-level change history generation tools.   
    \item We introduce a new algorithm and tool, \textit{HistoryFinder}, that addresses the limitations of the \textit{CodeShovel} tool.
    \item We conduct a thorough evaluation of the state-of-the-art method tracking tools against two accurate oracles to assess their relative performance in terms of precision, recall, F1 score, and runtime efficiency.
    \item We provide a user-friendly GUI-based tool that integrates all the state-of-the-art method-tracking tools (excluding \textit{IntelliJ}), allowing users to analyze and generate method histories for any arbitrary Java method from a \texttt{Git-based} project.
\end{itemize}


We publicly share the two oracles and the GUI-based history generation tool.\footnote{\textit{HistoryFinder}: \url{https://github.com/SQMLab/HistoryFinder}} While this GUI-based tool can be used by both practitioners and researchers, our shared repository also shows how to use the new \textit{HistoryFinder} tool from the command line for enabling research in \emph{mining software repositories}.
\section{Background and Related Work}
In this section, we discuss earlier research that leveraged code change history, especially at the method level, to analyze and build maintenance models. We then discuss research that focused on building history construction tools. 

\subsection{Research on Code Change History and Maintenance}
Software maintenance models commonly target the prediction of changes and bugs, as components that undergo frequent modifications or are tied to bug fixes are often harmful~\cite{chowdhury_good_2025, arvanitou_method_2017}. Early identification of such components can substantially lower future maintenance effort and cost~\cite{gunes_koru_identifying_2007, abbas_software_2020, chowdhury_good_2025}. Leveraging various code metrics, Koru \textit{et al.}~\cite{gunes_koru_identifying_2007} built tree-based models to forecast change-prone classes. In a similar vein, Romano \textit{et al.}~\cite{romano_using_2011} applied code metrics to predict change-prone \texttt{Java} interfaces. Khomh \textit{et al.}~\cite{khomh_exploratory_2009} demonstrated that certain types of code smells can also signal classes likely to change. Bug prediction models pursue a similar objective to \textit{change-proneness} prediction models: identifying code elements—such as classes or files—that are more likely to contain bugs in future versions~\cite{aleithan_explainable_2021, alsolai_predicting_2018, basili_validation_1996, gil_correlation_2017, zimmermann_predicting_2007, kamei_defect_2016}. These studies, however, focused on the class or file level, which has been reported as less useful by both practitioners and researchers~\cite{grund_codeshovel_2021, pascarella_performance_2020, giger_method-level_2012, chowdhury_method-level_2024}. For example, finding bugs in a class or file can be difficult due to their large size. Consequently, method-level change and bug prediction models have gained traction in recent years.

Method-level bug prediction has been studied by Giger \textit{et al.}~\cite{giger_method-level_2012}, Mo \textit{et al.}~\cite{mo_exploratory_2022}, Ferenc \textit{et al.}~\cite{ferenc_automatically_2020}, Shippey \textit{et al.}~\cite{shippey_so_2016}, Hata \textit{et al.}~\cite{hata_bug_2012}, Menzies \textit{et al.}~\cite{menzies_data_2007}, Pascarella \textit{et al.}~\cite{pascarella_performance_2020}, and Chowdhury \textit{et al.}~\cite{chowdhury_method-level_2024}. These studies heavily relied on constructing method-level change histories—either to build labeled datasets or to derive change metrics used as input features for machine learning models. For instance, Mo \textit{et al.}~\cite{mo_exploratory_2022} employed metrics such as the number of added and deleted lines within a method to predict its bug-proneness, a strategy also adopted by Pascarella \textit{et al.}~\cite{pascarella_performance_2020}. Chowdhury \textit{et al.}~\cite{chowdhury_method-level_2024} used method-level change history to identify bug-fixing commits and label methods as buggy or non-buggy. We argue that the accuracy of the tools used to construct method-level change histories is critical, as inaccuracies can lead to erroneous change metrics and incorrect bug labels.

\subsection{Research on History Construction}
Earlier studies on method-level tracking in \texttt{Git} used separate files to record method changes alongside \texttt{Git’s} file-level history. After extensive preprocessing of the entire codebase, this approach enables the use of \texttt{git log} to retrieve method histories. Hata \textit{et al.}~\cite{hata_historage_2011} introduced this technique with \textit{Historage}, which tracks changes to constructors, methods, and fields. However, Higo \textit{et al.}~\cite{higo_tracking_2020} found that \textit{Historage} struggles to track small methods when they are renamed or moved across files. To overcome these issues, they proposed \textit{FinerGit}, which represents \texttt{Java} methods with one token per line and applies heuristics for change comparison. Still, the preprocessing is time-intensive and incurs significant storage overhead—an issue common across similar tools~\cite{kim_when_2005,tu_integrated_2002,zimmermann_fine-grained_2006,hassan2004c}. Other tools~\cite{huang_cldiff_2018,falleri_fine-grained_2024} analyze changes between two versions but fall short of supporting origin analysis required by developers~\cite{grund_codeshovel_2021} and researchers~\cite{chowdhury_good_2025}. 

Recent studies have proposed solutions that eliminate the need for upfront repository preprocessing and avoid tracking methods at each \texttt{Git} commit~\cite{jodavi_accurate_2022, grund_codeshovel_2021}. Grund \textit{et al.}~\cite{grund_codeshovel_2021} introduced \textit{CodeShovel}, a state-of-the-art tool capable of recovering method histories despite various transformations. On a manually annotated oracle, \textit{CodeShovel} outperformed both research tools (e.g., \textit{FinerGit}) and industry tools (e.g., \textit{IntelliJ}, \textit{GitLog}). However, Jodavi \textit{et al.}~\cite{jodavi_accurate_2022} found that \textit{CodeShovel} struggles when methods undergo significant body changes, such as during moves, often failing to track them correctly or misattributing their origin.

They also noted inaccuracies in the original \emph{CodeShovel oracle} and revised it based on their interpretation to create a more reliable benchmark. They developed \textit{CodeTracker}, which employs \texttt{RefactoringMiner’s} advanced entity matching algorithm~\cite{tsantalis_refactoringminer_2022}. When evaluated with the new oracle, \textit{CodeTracker} showed significant improvements in precision and recall over \textit{CodeShovel}, but its reliance on \texttt{RefactoringMiner} introduced a performance cost, making it significantly slower.

\emph{\textbf{Motivation:}} Like its predecessor, the new \textit{CodeTracker} oracle remains subject to internal validity threats, as it relies entirely on researchers’ interpretations. A more systematic oracle construction process is needed for reliable evaluation of method history tools. Additionally, there is a need for a new algorithm that maintains high accuracy without sacrificing runtime performance, unlike \textit{CodeTracker}.
\section{Developing the New Oracles}
\label{sec:developing-oracles}
To evaluate method-tracking tools, it is essential to have reliable ground truths against which the correctness of the tools can be measured.
A completely manually constructed oracle remains vulnerable to internal validity threats due to its substantial dependence on human judgment. Manual efforts are prone to inconsistencies, oversight, and bias, particularly when dealing with complex software evolution patterns. Constructing such an oracle is also highly impractical for large-scale real-world codebases, which often consist of thousands of files and methods. During software evolution, methods may undergo renaming, move across files, or be refactored in tandem with other code changes, including modifications to their signatures and bodies. These transformations may occur within a single commit or span across multiple commits, making manual tracking extremely error-prone and difficult to scale. 

We, therefore, adopted a hybrid approach that combined automated analysis with human verification. We collected method histories from multiple state-of-the-art tools—each with distinct strengths and weaknesses—and manually validated each reported change to ensure high precision and recall. This union-based strategy was inspired by an observation in the original \textit{CodeShovel} paper: although \textit{CodeShovel} outperformed other tools, it occasionally missed method change commits that were correctly identified by other tools.

\begin{figure}[t]
\centering
\resizebox{\columnwidth}{!}{%
\begin{tikzpicture}[
  node distance=0.4cm and 0.6cm, 
  font=\scriptsize,              
  every node/.style={scale=0.9}  
]

\node (tool1) [tool] {\faTools\quad CodeShovel};
\node (tool2) [tool, below=of tool1] {\faTools\quad CodeTracker};
\node (tool4) [tool, below=of tool2] {\faLaptopCode\quad IntelliJ};
\node (tool5) [tool, below=of tool4] {\faGit\quad GitLineRange};
\node (tool6) [tool, below=of tool5] {\faGit\quad GitFuncName};

\node (mergepoint) [coordinate, right=1.8cm of tool4] {}; 

\node (aggregation) [process, right=2.8cm of tool4] {\faBezierCurve\quad Aggregate Method History};

\node (review) [manual, below=0.5cm of aggregation] {\faUsers\quad Manual Review};

\node (output) [output, below=0.5cm of review] {\faHistory\quad Refined Method History};

\foreach \i in {1,2,4,5,6} {
  \draw [thick] (tool\i.east) .. controls +(right:8mm) and +(left:8mm) .. (mergepoint.west);
}

\draw [arrow] (mergepoint) -- (aggregation.west);

\draw [arrow] (aggregation) -- (review);
\draw [arrow] (review) -- (output);

\begin{pgfonlayer}{background}
    \node [draw=gray, fill=gray!10, rounded corners, fit=(tool1)(tool6), inner sep=0.2cm, label=above:{\large\textbf{Collect Method History}}] {};
\end{pgfonlayer}

\end{tikzpicture}
} 
\caption{The process to generate accurate oracles. We first took the union of the change commits reported by all tools. The union of all commits was then manually verified by two authors to remove the false positives. The authors also added any missing commits they found that were not captured by any of the tools.}
\label{fig:oracle-construction-process}
\end{figure}
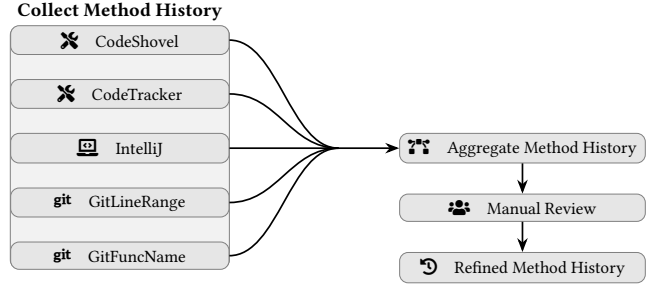

Each method’s history (i.e., the union of all detected method change commits) was independently reviewed by the first and second authors—both of whom have 8+ years of industry experience in software development and \texttt{Git-based} version control project management. They also identified and incorporated any missing commits that were not captured by the automated tools. They then discussed the few discrepancies via a Zoom meeting to reach a consensus. Figure~\ref{fig:oracle-construction-process} summarizes the oracle construction process. Next, we describe the two oracles we developed.

\subsection{Updating the CodeShovel Oracle}
\label{sec:updating-codeshovel-oracle}
Using the five tools \textit{CodeShovel}, \textit{CodeTracker}, \textit{IntelliJ}, \textit{GitLineRange}, and \textit{GitFuncName}), we collected the method change histories of the 200 methods used in the original CodeShovel study. For \textit{CodeShovel} and \textit{CodeTracker}, we utilized them as library dependencies, as they are publicly available on \texttt{GitHub}. For \textit{IntelliJ}, we manually imported each project, checked out the specific commits, located the target files and methods, and used the \texttt{Show History for Method} option to generate the method change histories. For the Git-based tools, we executed shell commands namely \texttt{git log <commit> --no-merges -L :<funcname>:<file>} and \texttt{git log <commit> --no-merges \allowbreak{}-L <start>,<end>:<file>} to extract method histories. Finally, we aggregated the method change histories from all five tools into a unified commit history, producing a union of commit histories of all tools. We then sorted the commit histories in reverse chronological order by their commit timestamp. Next, we manually inspect each method change commit to verify its logical validity by comparing it with its parent commit. This process is repeated iteratively for each subsequent commit until all commits have been thoroughly examined. At the end of this procedure, we obtained a complete and accurate method history oracle for the 200 methods. We refer to these 200 methods as the corrected \emph{CodeShovel oracle}. 

\subsection{Constructing the HistoryFinder Oracle}
\label{sec:history-finder-oracle}
We selected 20 additional open-source Java projects from \texttt{GitHub} to create the \emph{HistoryFinder oracle}. To ensure high-quality projects~\cite{kalliamvakou_promises_2014}, we selected popular projects based on a high number of stars and forks. We also ensured that each project must have at least 2,000 commits so that enough method change commits are found for the selected methods from these projects. 
Table \ref{tab:history-finder-repository-list} presents the list of selected projects along with their number of commits, Java files, methods, and their respective star and fork counts. From each project, we randomly selected 10 methods, recording their file names, method names, and start and end line numbers, resulting in a total of 200 methods. We then followed the same approach as described in section \ref{sec:updating-codeshovel-oracle} and generated the change history oracle for this new set of 200 methods.

\begin{table}[ht]
\centering
\caption{Overview of the 20 Java repositories used for the \emph{HistoryFinder oracle}, including commit counts, source files, methods, and popularity metrics (stars and forks).}
\label{tab:history-finder-repository-list}
\resizebox{\linewidth}{!}{%
\begin{tabular}{|c|c|c|c|c|c|c|c|}
\hline
\textbf{repository} & \textbf{\# commits} & \textbf{\# files} & \textbf{\# methods} & \textbf{\# stars} & \textbf{\# forks} \\ \hline
    auto & 2,570 & 317 & 4,211 & 10,441 & 1,202 \\
    cassandra & 31,705 & 4,926 & 76,475 & 8,913 & 3,641 \\
    dagger & 5,592 & 1,436 & 8,043 & 17,467 & 2,019 \\
    dubbo & 9,428 & 3,713 & 28,365 & 40,581 & 26,442 \\
    fastjson & 7,216 & 2,997 & 19,239 & 25,761 & 6,498 \\
    glide & 3,395 & 724 & 8,804 & 34,718 & 6,128 \\
    guice & 2,444 & 615 & 6,981 & 12,521 & 1,673 \\
    Hystrix & 2,354 & 415 & 6,436 & 24,160 & 4,708 \\
    j2objc & 6,314 & 5,119 & 106,062 & 5,998 & 972 \\
    jenkins & 37,342 & 1,779 & 21,660 & 23,363 & 8,834 \\
    jmeter & 32,202 & 1,526 & 16,138 & 8,468 & 2,116 \\
    kafka & 20,356 & 4,632 & 62,991 & 29,077 & 14,030 \\
    MPAndroidChart & 2,074 & 223 & 2,447 & 37,694 & 9,027 \\
    netty & 19,625 & 1,973 & 22,902 & 33,578 & 15,974 \\
    pulsar & 20,639 & 4,153 & 44,183 & 14,323 & 3,601 \\
    rocketmq & 8,918 & 1,581 & 17,155 & 21,361 & 11,727 \\
    RxJava & 7,698 & 1,884 & 33,994 & 47,951 & 7,602 \\
    shardingsphere & 43,769 & 7,765 & 34,921 & 20,035 & 6,775 \\
    tomcat & 77,289 & 2,705 & 42,021 & 7,619 & 5,062 \\
    zxing & 3,816 & 499 & 3,293 & 32,923 & 9,365 \\
\hline
\end{tabular}
}
\end{table}

\section{Developing the HistoryFinder Tool and Algorithm}
\label{sec:history-finder-algorithm}

The first, third, and fourth authors, each with significant industry and research experience, collaboratively designed the \emph{HistoryFinder} algorithm through multiple Zoom meetings. During these sessions, they analyzed the strengths and limitations of two existing approaches, \textit{CodeShovel} and \textit{CodeTracker}, which guided the design of their new algorithm. Their analysis revealed that while \textit{CodeTracker} is generally more accurate than \textit{CodeShovel}, it suffers from significantly slower runtime performance. This is primarily due to its dependence on RefactorMiner, whereas \textit{CodeShovel} uses a simpler, faster string-matching algorithm. To balance accuracy and efficiency, \textit{HistoryFinder} adopts \textit{CodeShovel’s} lightweight string-matching approach but improves upon it with a more precise method-tracking strategy.

\textit{HistoryFinder} also improved some of the technical weaknesses that \textit{CodeShovel} has. For example, during development, if code is merged without rebasing, conflicts can arise when a method is modified concurrently in different branches. The resolution of these conflicts may introduce changes to the method’s history that must be considered. \textit{CodeShovel}, however, excludes merge commits and ignores changes introduced during merges. In our approach, we include merge commits and compare the method’s state with each parent commit. This allows us to accurately detect any changes introduced from either parent branch during a merge process. Moreover, \textit{CodeShovel} does not account for formatting changes, annotations, or JavaDoc updates, whereas our proposed algorithm does. Another limitation of \textit{CodeShovel} is its case-insensitive string-matching algorithm. This approach can be problematic, especially when comparing quoted text within a method body, as it may overlook subtle but important differences that could influence the method's behavior.

\subsection{Overview of HistoryFinder}
The \textit{HistoryFinder} tool requires several key inputs to trace the history of a method. These include the \texttt{Git} repository URL or local directory, a specific commit hash (SHA), the file path of the method (including the file name), the name of the method, and the line number where the method declaration begins. The output is a sequential list of commit hashes, each accompanied by relevant metadata. This metadata provides granularity of changes, such as changes to the method signature, body, annotations, and identifies the contributor responsible for each modification. When executed via the command line, the output is formatted and saved as a \texttt{JSON} file.

We utilize the \texttt{JGit} library\footnote{https://github.com/eclipse-jgit/jgit last accessed: Jul 01, 2025}—a \texttt{Java}-based \texttt{Git} implementation—for repository cloning, commit traversal, and fine-grained file operations, including detection of added, deleted, moved, and modified files across revisions, and employ \texttt{JavaParser}\footnote{https://github.com/javaparser/javaparser last accessed: Jul 01, 2025} to generate an Abstract Syntax Tree (AST). Once the AST is constructed, it allows us to traverse through the list of methods in the file and extract specific components, such as the method signature, body, annotations, and JavaDoc. For method comparison, we use the Jaro-Winkler edit distance~\cite{winkler90}, as in \textit{CodeShovel}, to score similarity between method signatures and bodies. Although we also experimented with \texttt{SimHash}~\cite{charikar_similarity_2002} for similarity matching, we found that Jaro-Winkler consistently provides better recall than \texttt{SimHash}.

\begin{algorithm}
\caption{History Finder Algorithm}
\label{alg:algorithm-history-finder}
\begin{algorithmic}[1]
\Procedure{HistoryFinder}{$hash,file,name,line$} \Comment{$name$ and $line$ are the method's name and start line}

    \State $G \gets \textsc{GitLog}(hash, file)$ \Comment{Build file-level DAG}
    \State $s \gets \textsc{Create}(hash,file,name,line)$
    \State $Q \gets \{s\}$ \Comment{Initialize queue with the starting commit node}

    \State $visited \gets \{s\}$ \Comment{Track visited commit nodes}
    \State $H \gets \emptyset$ \Comment{Initialize with empty method change history}
    \While{$Q \neq \emptyset$} \Comment{Process all discovered commit nodes}
        \State $u \gets \textsc{Dequeue}(Q)$
        \State $um \gets \textsc{Find}(u.hash, u.file, u.name, u.line)$ 
        \State $p \gets \textsc{Parent}(G, u)$ \Comment{Get direct parent commit of $u$}
        \State $m \gets \textsc{FindBySignatureMatch}(p,file,um)$
        \State $altFile \gets \textsc{NIL}$ \Comment{Track method or file move}
         \If{$m == \text{NIL}$} \Comment{No match by signature}
            \State $m \gets \textsc{FindByBodyMatch}(p,file,um, 0.70)$
            \If{$m == \text{NIL}$} \Comment{No match by body similarity}
                \State $m,altFile \gets \textsc{AltFileMatch}(u.hash,p,um)$ 
            \EndIf
        \EndIf
        \If{$m \not= \text{NIL}$} \Comment{Method found}
            \If{$um.\text{text} \not= m.\text{text} \textbf{ or } altFile \not= \textsc{NIL}$}
                \State \textsc{Add}$(H, u)$ \Comment{Record change}
            \EndIf
            \If{$altFile \not= \textsc{NIL}$} \Comment{Method or file moved}
                \State $sH \gets \textsc{HistoryFinder}(m.hash,m.file,m.line)$ 
                \State \textsc{AddAll}$(H, sH)$ \Comment{Append sub-history}
            \EndIf
        \If{$m \not= \textsc{NIL} \textbf{ and } altFile == \textsc{NIL}$} 
            \ForAll{$v \in G.\text{Adj}[u]$} \Comment{Explore ancestors}
                \If{$v \notin visited$}
                    \State \textsc{Add}$(visited, v)$
                    \State $v.name \gets m.name$
                    \State $v.line \gets m.line$
                    \State \textsc{Enqueue}$(Q, v)$ \Comment{Add to queue}
                \EndIf
            \EndFor
        \EndIf
        \EndIf
    \EndWhile    
    \State \Return $H$
\EndProcedure
\end{algorithmic}
\end{algorithm}

\begin{algorithm}
\caption{Method Matching with Similarity Utility Procedures}
\label{alg:history-finder-similarity-matching}
\begin{algorithmic}[1]
\Procedure{FindBySignatureMatch}{$hash,file,tm$} \Comment{Match target method by exact signature}
    \State $methods \gets \textsc{method}(hash, file)$ \Comment{Parse methods from file}
    \ForAll{$m \in methods$} \Comment{Iterate through parsed methods}
        \If{$m.\text{signature} == tm.\text{signature}$} \Comment{Match signature}
            \State \Return $m$ \Comment{Return matched method}
        \EndIf
    \EndFor
    \State \Return $\text{NIL}$ \Comment{No match found}
\EndProcedure
\Procedure{FindByBodyMatch}{$hash,file, tm, theshold$} \Comment{Match target method by body similarity}
    \State $methods \gets \textsc{method}(hash, file)$ \Comment{Parse methods from file}
    \State $bm \gets \text{NIL}$ \Comment{Track best matching method}
    \State $bs \gets 0$ \Comment{Track highest similarity score}
    \ForAll{$m \in methods$} \Comment{Iterate through all methods}
        \State $score \gets \textsc{Simlarity}(m.text,sm.text)$ \Comment{Get similarity}
        \If{$score > theshold \textbf{ and } score > bs$}
            \State $bm \gets m$ \Comment{Track optimal matching method}
            \State $bs \gets score$ \Comment{Track optimal score}
        \EndIf
    \EndFor
    \State \Return $bm,bs$ \Comment{Return best match and score}
\EndProcedure
\Procedure{AltFileMatch}{$hash,parent\_hash,tm$} \Comment{Search for target method in other modified files}
    \State $changeFiles \gets \textsc{findChangeFiles}(hash,parent\_hash)$ \Comment{List modified files between commits}
    \State $bm \gets \text{NIL}$ \Comment{Track best match across files}
    \State $bs \gets 0$ \Comment{Track best score across files}
    \State $altFile \gets \textsc{NIL}$ \Comment{Store file where best match is found}
    \ForAll{$file \in changeFiles$} \Comment{Check each modified file}
        \State $m,score \gets \textsc{FindByBodyMatch}(parent\_hash,file,tm,0.75)$
        \If{$score > bs$} \Comment{Update best match if score improves}
            \State $bm \gets m$
            \State $bs \gets score$
            \State $altFile \gets file$
        \EndIf
    \EndFor
    \State \Return $bm,altFile$  \Comment{Return best match and its file}
\EndProcedure
\end{algorithmic}
\end{algorithm}

All commits leading up to a specific point in a \texttt{Git} project form a Directed Acyclic Graph (DAG)~\cite{geisshirt_git_2018}, representing the project's complete commit history to that point. To identify all commits that modified a method up to a given version (e.g., the starting commit), we traverse this DAG backward—from the target commit to its ancestors—until we reach the commit where the method was originally introduced. During this traversal, if a commit has a single parent, we compare the method in the current commit with its counterpart in the parent commit to detect changes. In the case of merge commits with multiple parents, the method may exist in any of the parent commits and must be analyzed accordingly. Given that a project may contain thousands of commits (as shown in Table~\ref{tab:history-finder-repository-list}), traversing the entire DAG can be time-consuming. To optimize this process, we leverage the \texttt{git log <commit> <file>} command, which filters the commit history to only those that modified the file containing the target method. This significantly reduces the search space, allowing us to traverse a lightweight, file-specific DAG to detect method changes more efficiently. However, if we discover that the method has been moved to a different file or the file itself has been renamed, we must reconstruct a new DAG from that point onward using the new file path.

\subsection{The HistoryFinder Algorithm}
Algorithm~\ref{alg:algorithm-history-finder} outlines how the file-specific DAG is constructed and traversed to detect the change history of a method. The procedure \textit{HistoryFinder} is initially invoked with the starting commit, file path, method name, and its starting line number. It may be invoked again if the method is detected to have moved to a different file or if the file itself has been renamed. To build the DAG, we first call the utility procedure \textit{GITLOG} (line 2), which internally executes the \texttt{git log <commit> <file>} command. This retrieves only the commits that have modified the target file containing the method. Using this filtered commits, we construct a directed acyclic graph \( G \), where each node represents a commit that changed the file. In the DAG, an edge is drawn from node $u$ to node $v$ if the file is modified in commit $u$, followed by in commit $v$, establishing a direct connection between them. Once the DAG is constructed, we traverse it using the Breadth-First Search (BFS) algorithm~\cite{cormen_introduction_2022} to systematically explore all relevant commits.

Queue \( Q \) keeps track of all discovered commit nodes that need to be explored. To unambiguously locate the target method in a commit, each commit node is represented by a structure containing the commit hash, file path, method name, and line number, created by invoking \texttt{Create} (line 3 for the initial node) procedure. The queue \( Q \) is initialized with this starting node (line 4), which is also the only entry in the set \texttt{visited} to mark it as explored. Whenever a change to the method is detected, it is recorded in the history list \( H \), which is initially empty (line 6). The \texttt{while} loop iterate through all discovered nodes. In each iteration, the first unexplored commit node \( u \) is dequeued (line 8). We locate the method already found in commit node \( u \) by calling the procedure \texttt{Find} (line 9), which takes the commit hash, file, method name, and line number as parameters, accessed using dot notation. Next, we retrieve the parent node \( p \) of \( u \) (line 10) where method might be found. From lines 11 to 18, we check whether the method exists in direct parent commit \( p \), accounting for possible modifications. 

We first try to locate the method by matching its signature (line 11), including the method name and parameters, but excluding the return type. If the signature match fails, we attempt to locate the method by computing the textual similarity of its body (line 14), considering a match valid if the similarity is at least 70\%. If the method still cannot be found, we broaden the search to other modified files by checking for body similarity across them (line 16), raising the similarity threshold to 75\% in this case. If the method is found in a different file—due to a method move or a file rename/move—we record this in the variable \texttt{altFile}. If the method is found (line 19), we then check whether the methods (from the current and parent commits) are identical, including their signatures, JavaDocs, annotations, or whether the file has been renamed or moved (in which case \texttt{altFile} should not be \texttt{NIL}) (line 20). If any change is detected—either in the method or in the file name or path—we record this change by invoking the \texttt{Add} procedure (line 21). 

If the method is found in another file, we must reconstruct a new DAG for that file. To do so, we recursively call the \texttt{HistoryFinder} procedure (line 24), which returns a sub-history of changes. This sub-history is then appended to the history list \( H \) (line 25). If the method is instead found within the same file (line 27), we proceed to process all immediate ancestor commit nodes. In this case, we add (line 30) and enqueue (line 33) all adjacent ancestor commit nodes that have not yet been visited. We also update the method name (line 31) and line number (line 32) to reflect the newly found version, so that it can be retrieved accurately when this node is dequeued later. Once all nodes have been processed, we obtain a sequence of commit nodes capturing the method’s change history in \(H\). Additionally, we add the furthest node dequeued from the queue $Q$ as the introduction commit, since this is the earliest known instance of the method.

Algorithm~\ref{alg:history-finder-similarity-matching} defines the utility procedures used to compare and match methods based on similarity for identifying the target method. The procedure \texttt{FindBySignatureMatch} attempts to locate a target method \texttt{tm} within a Java file at a specific commit. Line 2 parses all methods in the file using \texttt{JavaParser}, and lines 3 to 7 iterate over these methods, comparing each one’s signature with that of \texttt{tm}. If a matching signature is found, the corresponding method is returned (line 5); otherwise, \texttt{NIL} is returned (line 8). The next procedure, \texttt{FindByBodyMatch} (line 10), identifies a method based on the highest body similarity with the target method \texttt{tm}, provided it exceeds a specified similarity threshold. A threshold of 70\% is used when comparing methods within the same file, and 75\% when comparing across different files. Line 11 parses all Java methods in the file provided as input. The variable \texttt{bm} (line 12) tracks the best matching method, while \texttt{bs} (line 13) stores the corresponding similarity score. Line 14 iterates over all parsed methods, and the \texttt{Similarity} procedure (line 15) computes the similarity between the current method and the target method using the Jaro-Winkler algorithm~\cite{winkler90}. If the computed similarity exceeds the threshold and is greater than the current best score \texttt{bs}, both the score and the best matching method are updated (lines 16–19). Finally, the best matching method along with its similarity score is returned (line 21). 

The procedure \texttt{AltFileMatch} attempts to locate the target method in all files that were modified in a given commit. It begins by listing all modified files using the \texttt{FindChangedFiles} procedure (line 24). Lines 25 to 27 initialize variables to track the best matching method, its similarity score, and the corresponding file path. Each modified file is then iterated over (line 28), and the best matching method within each file is identified (line 29). If a method with a higher similarity score is found, the method information is updated accordingly (lines 30–34). Finally, the method with the highest similarity score and its associated metadata are returned (line 36). Similar to the \textit{CodeShovel} algorithm~\cite{grund_codeshovel_2021}, we used the first 100 methods from the corrected \textit{CodeShovel} oracle to train our algorithm and determine the optimal thresholds for various similarity scores.
\section{Analysis and Results}
\label{sec:result-and-analysis}
In this section, we compare the corrected \emph{CodeShovel oracle} with the original \emph{CodeShovel and CodeTracker oracles}. We then evaluate the accuracy and runtime efficiency of all method change history generation tools using the corrected \textit{CodeShovel} and the newly developed \emph{HistoryFinder oracles}. In particular, we answer the following three research questions.
\begin{itemize}
    \item \textbf{RQ1}: To what extent does the corrected \textit{CodeShovel} oracle differ from the original oracles? 
    \item \textbf{RQ2}: Which history generation tool exhibits the highest accuracy when evaluated with the two new oracles?
    \item \textbf{RQ3}: Which history generation tool exhibits the best runtime performance? 
\end{itemize}
\subsection{RQ1: Difference with the corrected CodeShovel oracle}


After correcting the method histories of the 200 methods in both the \textit{CodeShovel} and \textit{CodeTracker} oracles, we compared our new systematically constructed oracle—\emph{the corrected CodeShovel oracle}—with the original \textit{CodeShovel} and \emph{CodeTracker oracles} and observed significant updates. For example, compared to the original \textit{CodeShovel} oracle, the histories of 40 methods were \emph{modified}, with 59 commits \emph{added} and 186 \emph{removed}. These changes exclude annotations, JavaDocs, and formatting—omitted by the original \textit{CodeShovel} and \textit{CodeTracker} oracles, though \textit{CodeTracker} did include annotations. When these types of changes were included, 142 method histories were affected, and 429 commits were \emph{added}. Significant changes were also observed when compared with the \emph{CodeTracker oracle}. Table~\ref{tab:oracle-update-statistics} provides a detailed breakdown of the changes. These changes suggest that the previous evaluations of method tracing tools (e.g., \textit{CodeShovel} and \textit{CodeTracker}) may not reflect the true precision and recall of these tools in tracing a method's change history.

\begin{table}[!ht]
\centering
\caption{Comparison of the corrected oracle with the original oracles. We show how many methods were \emph{changed}, commits were \emph{added} and \emph{removed}. A method is considered \emph{changed} if the new oracle has a different set of commits than the original. Similarly, a commit is considered \emph{added} if it was not present in the original oracles, and \emph{removed} if it was inaccurately included. For example, the last column shows that the \emph{CodeTracker oracle} has fewer inaccurate commits than the original \emph{CodeShovel oracle}. In contrast, it missed many commits that should have been recorded as change commits for those 200 methods, as shown in the \emph{\# Added Commits} column. As the corrected oracle includes changes related to annotations, JavaDocs, and formatting, we present the differences both including (\emph{Incl.}) and excluding (\emph{Excl.}) those changes.}
\label{tab:oracle-update-statistics}
\resizebox{\linewidth}{!}{%
\begin{tabular}{|c|cc|cc|c|}
\hline
\multirow{2}{*}{\textbf{Oracle}} & \multicolumn{2}{|c|}{\textbf{\# Changed Methods}} & \multicolumn{2}{|c|}{\textbf{\# Added Commits}} & \multirow{2}{*}{\textbf{\# Removed Commits}} \\ \cline{2-5}
 & Incl. & Excl. & Incl. & Excl. & \\ \hline
CodeShovel & 142 & 40 &  429 & 59 & 186 \\
CodeTracker & 133 & 43 &  445 & 134 & 67 \\
\hline
\end{tabular}
}
\vspace{2mm}
\end{table}

To illustrate inaccuracies in both the \textit{CodeShovel} and \textit{CodeTracker} oracles, we highlight several cases. One such example is the \texttt{fireErrors} method from the \texttt{checkstyle} project. Both the \textit{CodeShovel}\footnote{\href{https://github.com/SQMLab/HistoryFinder/blob/main/oracle/codeshovel-oracle/checkstyle-Checker-fireErrors.json}{https://github.com/SQMLab/HistoryFinder/blob/main/oracle/codeshovel-oracle/checkstyle-Checker-fireErrors.json}} and \textit{CodeTracker}\footnote{\href{https://github.com/SQMLab/HistoryFinder/blob/main/oracle/codetracker-oracle/checkstyle-Checker-fireErrors.json}{https://github.com/SQMLab/HistoryFinder/blob/main/oracle/codetracker-oracle/checkstyle-Checker-fireErrors.json}} oracles record this method’s history beginning at commit \texttt{119fd4}, where it was located in \path{src/main/java/com/puppycrawl/tools/checkstyle/Checker.java}. However, both oracles incorrectly identified commit \texttt{0e3fe} as the method’s introduction, whereas it was actually introduced earlier in commit \texttt{0fd69}. At that point, the method name was \texttt{displayErrors} and was later renamed to \texttt{fireErrors}. This error was recently corrected in the \emph{CodeTracker} oracle; however, our evaluation was conducted using the previously published version in which the issue remained unresolved.

In another case, while tracking overloaded methods, the \emph{CodeShovel oracle} failed to stop at the point where the overload was introduced and instead switched (\texttt{createPattern} method\footnote{\CodeShovelOverloadedCreatePatternFootnote}) to tracing the original method, resulting in incorrect method changes in the oracle. Another challenge arises when handling merge commits. While the \emph{CodeShovel oracle} completely ignored changes that occurred in merge commits, the \emph{CodeTracker oracle} occasionally added incorrect method changes. An example of this issue occurs in the \texttt{runChild} method of the \texttt{junit4} project. Although the diff\footnote{\CodeTrackerMergeCommitRunChildFootnote} appears to show a method change, it is incorrect—the commit has two parents,\footnote{\RunChildMergeCommitParentFootnote} and the comparison was made against the wrong parent.\footnote{\RunChildWrongParentFootnote} In reality, the code was introduced from the other branch\footnote{\RunChildCorrectParentFootnote} without any modification.

\begin{tcolorbox}
\textbf{\underline{RQ1 summary}:} 
Our systematic approach to correcting the existing \emph{CodeShovel oracle} reveals that both the \emph{CodeShovel} and \emph{CodeTracker oracles} contain inaccuracies in the constructed change histories of methods. In some cases, they include incorrect commits; in others, they miss valid change commits. The updated, more accurate oracle can serve as a valuable resource for effectively evaluating both existing and future method-level change history generation tools.
\end{tcolorbox}

\subsection{RQ2: Accuracy of the change history generation tools}
\label{sec:rq2-accuracy}

We now evaluate the accuracy of six method change history generation tools, including the newly proposed \textit{HistoryFinder}. This evaluation uses the corrected \emph{CodeShovel oracle} (200 methods) and the new \emph{HistoryFinder oracle} (another 200 methods). The \textit{CodeShovel} oracle is divided into two parts: \emph{CodeShovel Training} and \emph{Testing} oracles. This distinction is important, as both the \textit{CodeShovel} and \textit{HistoryFinder} algorithms were trained and tuned using the training portion of the \emph{CodeShovel oracle}. We ran the three research-based tools as Java libraries to extract method histories and compute accuracy metrics. For Git-based tools, we used the command \texttt{git log <commit> --no-merges -L :<funcname>:<file>} to generate method histories for \textit{GitFuncName}, which tracks changes by function name, and \texttt{git log <commit> --no-merges\allowbreak{} -L <start>,\allowbreak{}<end>:<file>}  for \textit{GitLineRange}, which tracks changes by line range. In contrast, for \textit{IntelliJ}, we used IntelliJ IDEA 2024.3 Community Edition to manually generate method histories (\texttt{Right-click} $\rightarrow$ \texttt{Git} $\rightarrow$ \texttt{Show History for Selection}), as no automated approach exists for running IntelliJ's method tracking feature. 

We calculated the precision, recall, and F1 score for all the tools against the three oracles. Since tools differ in the types of changes they detect (e.g., \textit{CodeShovel} ignores annotations, \textit{CodeTracker} ignores formatting), we tailored the expected commit sets to match each tool’s capabilities, ensuring a fair and meaningful comparison. We evaluate the tools' accuracy at two granularities: \emph{commit-level} and \emph{method-level}. At the \emph{commit-level}, we compute precision, recall, and F1 scores by aggregating true positives (TP), false positives (FP), and false negatives (FN) across each oracle, based on the total expected commits. At the \emph{method-level}, metrics are computed per method and averaged across each oracle.

\begin{table*}[t]
\centering
\caption{Precision, recall, and F1 score for method history tracking tools across three oracles. 
\textbf{TP} (True Positives) indicates correctly identified change commits; 
\textbf{FP} (False Positives) refers to incorrectly included commits; 
\textbf{FN} (False Negatives) denotes missed commits that should have been identified. Precision is the proportion of correct results among all predicted changes, recall is the proportion of actual changes correctly identified, and F1 score is the harmonic mean of precision and recall. \textbf{Bolded values indicate the best performance in each metric within a oracle.}
}
\label{tab:method-tracking-score}
\resizebox{\textwidth}{!}{%
\begin{tabular}{|c|c|ccc|ccc|ccc|}
\hline
\multirow{2}{*}{\textbf{Oracle}} & \multirow{2}{*}{\textbf{Tool}} & \multicolumn{6}{c|}{\textbf{\emph{commit-level}}} & \multicolumn{3}{c|}{\textbf{\emph{method-level}}} \\
\cline{3-11}
 &  & \textbf{TP} & \textbf{FP} & \textbf{FN} & \textbf{Precision} & \textbf{Recall} & \textbf{F1} & \textbf{Precision} & \textbf{Recall} & \textbf{F1} \\
\hline
\multirow{6}{*}{CodeShovel Training}
    & CodeShovel & 2813 & 158 & 28 & 94.68 & \textbf{99.01} & 96.80 & 94.79 & \textbf{99.31} & 95.86 \\
    & CodeTracker & 2804 & 32 & 76 & 98.87 & 97.36 & 98.11 & 98.99 & 96.98 & 97.52 \\
    & HistoryFinder & 3047 & 6 & 67 & \textbf{99.80} & 97.85 & \textbf{98.82} & \textbf{99.68} & 98.15 & \textbf{98.53} \\
    & IntelliJ & 2146 & 402 & 919 & 84.22 & 70.02 & 76.47 & 85.16 & 75.08 & 73.40 \\
    & GitLineRange & 2197 & 1135 & 787 & 65.94 & 73.63 & 69.57 & 77.94 & 76.32 & 68.29 \\
    & GitFuncName & 1749 & 1166 & 1261 & 60.00 & 58.11 & 59.04 & 72.93 & 62.03 & 54.88 \\
\hline
\multirow{6}{*}{CodeShovel Testing}
    & CodeShovel & 812 & 25 & 38 & 97.01 & 95.53 & 96.27 & 97.40 & 95.70 & 95.54 \\
    & CodeTracker & 811 & 35 & 63 & 95.86 & 92.79 & 94.30 & 97.59 & 94.36 & 94.76 \\
    & HistoryFinder & 938 & 10 & 7 & \textbf{98.95} & \textbf{99.26} & \textbf{99.10} & \textbf{99.15} & \textbf{99.24} & \textbf{99.15} \\
    & IntelliJ & 685 & 239 & 249 & 74.13 & 73.34 & 73.74 & 85.14 & 72.58 & 73.78 \\
    & GitLineRange & 609 & 288 & 279 & 67.89 & 68.58 & 68.24 & 87.48 & 70.02 & 72.31 \\
    & GitFuncName & 579 & 393 & 331 & 59.57 & 63.63 & 61.53 & 77.85 & 65.34 & 65.35 \\
\hline
\multirow{6}{*}{HistoryFinder}
    & CodeShovel & 2095 & 147 & 394 & 93.44 & 84.17 & 88.56 & 95.68 & 86.45 & 85.82 \\
    & CodeTracker & 2354 & 88 & 168 & \textbf{96.40} & 93.34 & 94.84 & 94.43 & 91.24 & 94.14 \\
    & HistoryFinder & 2835 & 106 & 48 & \textbf{96.40} & \textbf{98.34} & \textbf{97.36} & \textbf{96.49} & \textbf{96.81} & \textbf{97.00} \\
    & IntelliJ & 2185 & 1530 & 637 & 58.82 & 77.43 & 66.85 & 75.99 & 81.38 & 75.80 \\
    & GitLineRange & 2032 & 2112 & 616 & 49.03 & 76.74 & 59.84 & 77.64 & 82.98 & 75.50 \\
    & GitFuncName & 1858 & 2638 & 835 & 41.33 & 68.99 & 51.69 & 65.08 & 72.67 & 59.50 \\
\hline
\end{tabular}
}
\end{table*}

Table~\ref{tab:method-tracking-score} shows that all research-based tools—\textit{CodeShovel}, \textit{CodeTracker}, and \textit{HistoryFinder}—significantly outperform \textit{IntelliJ} and \texttt{Git-based} tools. Overall, \textit{HistoryFinder} achieves highest F1 score. In terms of precision, \textit{HistoryFinder} and \textit{CodeTracker} slightly outperform \textit{CodeShovel}. For recall, \textit{CodeShovel} performs best on its own training oracle; however, on the other two oracles, \textit{HistoryFinder} achieves the highest recall. \textit{CodeShovel’s} notably lower recall on the \textit{HistoryFinder} oracle is primarily due to its failure to generate complete history for several methods in the oracle. For example, it was unable to produce complete histories for all 10 methods from the Cassandra project. Additionally, it failed entirely for two methods in the Fastjson project due to \texttt{JavaParser} errors that prevented successful file parsing. Overall, across all three oracles, \textit{HistoryFinder} demonstrates the most consistent and superior performance. Results at the \emph{method-level} scores closely correlate with \emph{commit-level} scores, with minor variations. For instance, \textit{CodeShovel's} precision against the \emph{HistoryFinder oracle} is approximately 2\% higher at the \emph{method-level} compared to the \emph{commit-level}. 

\begin{tcolorbox}
\textbf{\underline{RQ2 summary}:} 
Our evaluation shows that the \textit{HistoryFinder} algorithm consistently achieves the highest precision and F1 score in change history detection. Between \textit{CodeShovel} and \textit{CodeTracker}, performance varies depending on the scenario—each tool outperforms the other in different contexts. Notably, \textit{CodeShovel's} recall drops significantly on the entirely new oracle of 200 methods, mainly because it fails to generate any history for a subset of methods. Given these findings, if a researcher or practitioner prioritizes accuracy when selecting a method history generation tool, \textit{HistoryFinder} is the most reliable choice.
\end{tcolorbox}

\subsection{RQ3: Runtime performance of the change history generation tools}
\label{sec:rq3-runtime}

We evaluated the tools on a machine equipped with \emph{32GB of RAM, an Intel(R) Core(TM) i7-8700 CPU @ 3.20GHz with 12 cores, running Ubuntu 22.04 LTS and Java 21 x86}. Execution time for \textit{IntelliJ} was not recorded since method histories were generated manually using \texttt{IntelliJ IDEA}, making it infeasible to measure the time taken. Prior to recording execution time, all repositories were cloned to a local hard disk. Initially, we integrated the JAR dependencies of the research-based tools into a single Java-based project and recorded their execution times for each method history. However, we observed that caching of parsed Java files across multiple method history runs—particularly in \textit{CodeShovel}—significantly reduced the runtime. While this optimization benefits batch processing, it does not yield deterministic or representative runtimes for tracing individual methods. To ensure a fair comparison, we instead executed each tool independently for each method history using the \texttt{Java -jar} command with the necessary arguments. This approach ensured that each method history was processed in isolation within a fresh environment, by launching a separate \texttt{JAR} process for each run and terminating it upon completion. 

\begin{figure*}[t]
  \includegraphics[width=\linewidth]{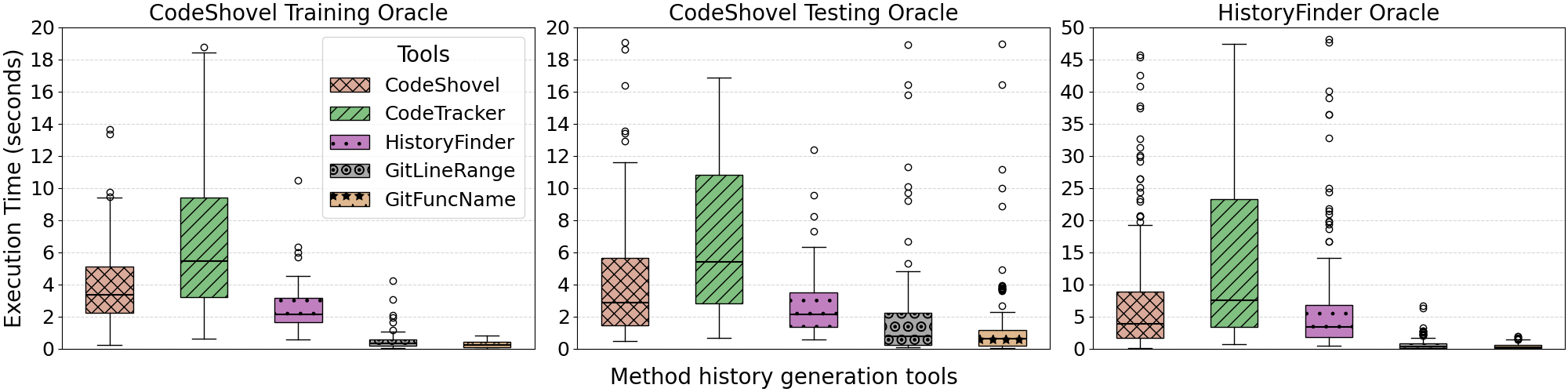}
\caption{Comparison of execution times across three oracles for five tools. To enhance readability, the y-axis is limited to 20 seconds for the first two subplots (\emph{CodeShovel} oracles) and 50 seconds for the third subplot (\emph{HistoryFinder} oracle).} 
\label{fig:execution-time-box-plot}  
\end{figure*}

Figure~\ref{fig:execution-time-box-plot} shows the distribution of execution times for all five tools across the three oracles, while Table~\ref{tab:execution-time-statistic} presents the mean, median, minimum, and maximum execution times. Excluding outliers, \textit{HistoryFinder} demonstrates the lowest execution times among the three research-based tools. In general, \textit{CodeTracker} is the slowest in constructing method change history. Across all tools and oracles, the minimum execution time consistently remained below one second. However, for the research-based tools, although the mean and median execution times are slightly higher in the \emph{HistoryFinder} oracle compared to the \emph{CodeShovel} oracle, the difference in maximum execution time between these two oracles is unexpectedly large.

To understand the factors contributing to the increased runtime in the \textit{HistoryFinder} oracles, we manually examined cases with unusually high execution times at both the individual method and project levels. We found that prolonged runtimes were often linked to a high frequency of changes in the file containing the target method, regardless of whether the method itself was modified. In such scenarios, more Java files needed to be parsed, and parsing using the \texttt{JavaParser} was particularly time-consuming. Our profiling of the tools revealed that, for methods with high execution times, over 80\% of the total runtime was typically spent on parsing alone, thereby significantly increasing the computational overhead. Repositories like Kafka, Cassandra, Jenkins, and Pulsar frequently exhibited these patterns and are more prevalent in the \emph{HistoryFinder} oracle compared to the \emph{CodeShovel} oracle. 

\textit{CodeTracker} exhibited significantly high execution times when a method or its enclosing file was moved in a commit that also modified many other files—a pattern observed in repositories such as Netty and ShardingSphere. In contrast, \textit{CodeShovel}'s maximum execution time on the \textit{HistoryFinder} oracle was much lower than that of \textit{CodeTracker} and \textit{HistoryFinder}. However, this lower runtime is primarily due to \textit{CodeShovel} failing to generate complete histories for many complex methods with extensive change histories—cases where both \textit{CodeTracker} and \textit{HistoryFinder} succeeded.


The \texttt{Git-based} tools, in contrast, did not exhibit high execution times in the \textit{HistoryFinder} oracle. Their maximum execution times were also significantly lower than those of the research-based tools. This lower runtime is primarily due to two factors: (i) \texttt{Git-based} tools operate using \texttt{git log} without relying on language-specific parsers. Instead, they track line-level changes, making their execution comparatively lightweight; and (ii) similar to \textit{CodeShovel}, these tools often failed to recover the complete change histories of complex methods. As a result, their efficiency comes at the cost of significantly lower precision and recall compared to their research-based counterparts, as also reflected in in Table~\ref{tab:method-tracking-score}.


\begin{table}[ht]

\centering
\caption{Mean, median, minimum, and maximum execution times (in seconds) for five tools. 
\textbf{Bold} values indicate the best (lowest) execution time for each metric within a oracle for a set of tools. 
\textit{IntelliJ} is excluded because method histories for this tool were generated manually using the \texttt{IDEA} interface.}
\resizebox{\linewidth}{!}{%
\begin{tabular}{|c|c|c|c|c|c|c|c|}
\hline
\textbf{Oracle} & \textbf{Tool} & \textbf{Mean} & \textbf{Median} & \textbf{Min} & \textbf{Max} \\ \hline
\multirow{5}{*}{\centering{CodeShovel Training}}
    & CodeShovel & 3.93 & 3.38 & \textbf{0.22} & 13.63 \\
    & CodeTracker & 8.71 & 5.45 & 0.64 & 76.80 \\
    & HistoryFinder & \textbf{2.51} & \textbf{2.15} & 0.59 & \textbf{10.48} \\
    \cline{2-6}
    & GitLineRange & 0.50 & 0.35 & 0.07 & 4.23 \\
    & GitFuncName & \textbf{0.26} & \textbf{0.24} & \textbf{0.02} & \textbf{0.83} \\
\hline
\multirow{5}{*}{\centering{CodeShovel Testing}}
    & CodeShovel & 4.55 & 2.86 & \textbf{0.49} & 26.08 \\
    & CodeTracker & 9.29 & 5.42 & 0.68 & 63.34 \\
    & HistoryFinder & \textbf{2.85} & \textbf{2.15} & 0.57 & \textbf{20.78} \\
    \cline{2-6}
    & GitLineRange & 2.13 & 0.80 & 0.11 & \textbf{18.93} \\
    & GitFuncName & \textbf{1.57} & \textbf{0.61} & \textbf{0.07} & 18.94 \\
\hline
\multirow{5}{*}{\centering{HistoryFinder}}
    & CodeShovel & 7.74 & 3.88 & \textbf{0.15} & \textbf{74.49} \\
    & CodeTracker & 30.41 & 7.56 & 0.76 & 693.66 \\
    & HistoryFinder & \textbf{7.32} & \textbf{3.47} & 0.49 & 106.64 \\
    \cline{2-6}
    & GitLineRange & 0.69 & 0.38 & 0.04 & 6.66 \\
    & GitFuncName & \textbf{0.41} & \textbf{0.21} & \textbf{0.03} & \textbf{1.96} \\
\hline

\end{tabular}
}
\label{tab:execution-time-statistic}
\end{table}

\begin{tcolorbox}
\textbf{\underline{RQ3 summary}:} 
\texttt{Git-based} tools demonstrated the lowest execution times due to their parser-free approach, but this came at the cost of reduced accuracy. Among the research-based tools, \textit{HistoryFinder} provides the best balance between execution time and accuracy, achieving the highest precision and recall along with the lowest mean and median execution times across oracles. In contrast, \textit{CodeTracker}, the second most accurate tool in history tracing, showed the highest and most variable runtimes, especially for complex histories involving large-scale refactorings. Based on these findings, \textit{HistoryFinder} emerges as the most reliable choice for researchers and practitioners who prioritize both accuracy and performance when selecting a method history generation tool.
\end{tcolorbox}

\section{Discussion}
\label{sec:discussion}
Due to its practical importance, method-level maintenance research has gained significant traction in recent years, making accurate method-level change history generation increasingly critical. The goal of this work was to assess whether existing tools for generating method-level change histories produce accurate and complete results when evaluated against rigorously constructed oracles, and to identify which tool offers the best balance between accuracy and runtime efficiency. Our study found that earlier evaluations based on the original \emph{CodeShovel} and \emph{CodeTracker} oracles were inaccurate, undermining the validity of prior assessments of tool performance (\textbf{RQ1}). To address this, we systematically constructed two new oracles—the corrected \emph{CodeShovel} oracle and the new \emph{HistoryFinder} oracle—offering a more reliable benchmark that combines automated detection with expert validation.

To further address the limitations observed in both \textit{CodeShovel} and \textit{CodeTracker} tools, we also developed a new algorithm and tool, \textit{HistoryFinder}, designed to improve both the accuracy and completeness of method change history tracking while maintaining competitive runtime performance. When evaluated against the newly constructed oracles, \textit{HistoryFinder} consistently achieved the highest precision, recall, and F1 scores  (\textbf{RQ2}) and demonstrated the most robust performance across all oracles, including those involving challenging repositories with extensive changes to the file containing the target method (\textbf{RQ3}). Although the \texttt{Git-based} tools delivered faster execution times, this efficiency came at the cost of substantially lower accuracy, making \textit{HistoryFinder} the most reliable option for users prioritizing both accuracy and reasonable runtime. 
Our runtime analysis also revealed that repositories with frequent file modifications result in longer processing times for tools that rely on \texttt{JavaParser}.

We hope that both the research and the practitioners communities can be benefited by our study. Researchers can leverage these oracles to train and evaluate new algorithms or improve existing method history generation tools, especially in handling complex scenarios—such as frequent file changes, large-scale refactorings that involve method or file moves, and tracking changes in annotations or JavaDocs associated with methods. 
To enhance accessibility and practical applicability, we provide a graphical user interface (GUI) that allows both researchers and practitioners to interactively run multiple tools—including \textit{CodeShovel}, \textit{CodeTracker}, \textit{HistoryFinder}, \textit{GitLineRange}, and \textit{GitFuncName}—on their own selected Java methods. Users can independently compare tool outputs and choose the one best suited to their needs, whether they prioritize accuracy or runtime performance. This tool can also be integrated into a developer’s daily workflow. 

In addition to the GUI, we provide support for a command-line interface (CLI) that can be used to track change histories of millions of methods for conducting MSR-based research. We also share our tool as a Java library for seamless integration into larger software engineering pipelines. The complete setup and usage instructions are provided with our shared repository.\footnote{\url{https://github.com/SQMLab/HistoryFinder}} 

\subsection{Threats to Validity}
\textbf{Internal Validity.}
Although we leveraged multiple state-of-the-art history generation tools to construct the two oracles, the validation of change commits still involved human evaluators. To mitigate this threat, two of the authors—each with over eight years of industry experience—independently conducted the validations. Nonetheless, we acknowledge that this threat could be further reduced by involving additional evaluators.

Similar to the \textit{CodeShovel} and \textit{CodeTracker} tools, a key limitation of the \textit{HistoryFinder} tool is its reliance on \texttt{JavaParser}, which may fail when processing syntactically incorrect files—often introduced by inadvertent changes during development.


\textbf{External Validity.}
Both the \emph{CodeShovel} and \emph{HistoryFinder} oracles were constructed using only open-source repositories, which may limit the generalizability of our findings. It remains uncertain whether the tools’ performance and accuracy would translate to closed-source or industrial settings, where code characteristics and commit practices may differ. While the \textit{HistoryFinder} tool can be extended to support other programming languages by incorporating language-specific parsers, it is unclear whether the similarity thresholds we trained for identifying method matches would remain effective outside of Java.

\section{Conclusion and Future Work}
\label{sec:conclusion}
In this study, we revisited the evaluation of \emph{method-level} change history generation tools by introducing two rigorously constructed oracles. These benchmarks address key limitations in prior oracles and enable more reliable assessments of tools' performance. Our empirical analysis shows that the newly developed \textit{HistoryFinder} tool consistently outperforms existing research-based tools such as \textit{CodeShovel} and \textit{CodeTracker} in terms of precision, recall, F1 score and runtime performance. Moreover, we provide a publicly accessible, user-friendly web interface and a command-line interface that allow researchers and practitioners to evaluate tools on their own method samples, making our findings immediately actionable in both academic and industrial settings.

For future work, we plan to extend \textit{HistoryFinder} to support programming languages beyond Java. We also aim to explore alternative parsing strategies—potentially language-independent solutions—as the current language-specific parser, \texttt{JavaParser}, has proven to be a performance bottleneck. Additionally, we intend to investigate alternative algorithms for detecting method similarity, including those based on machine learning techniques.

\balance
\bibliographystyle{ACM-Reference-Format}
\bibliography{references_saved, references}




\end{document}